\documentclass[epj,final]{svjour}
\usepackage{amsmath,amssymb,graphicx}
\usepackage[numbers,sort&compress]{natbib}
\newcommand{\avg}[1]{\langle #1\rangle}
\newcommand{\abs}[1]{\lvert #1\rvert}
\newcommand{\req}[1]{(\ref{#1})}
\newcommand{\NN}{\mathcal{N}}
\newcommand{\UU}{\mathcal{U}}
\newcommand{\dd}{\mathrm{d}}
\newcommand{\ee}{\mathrm{e}}
\newcommand{\erf}{\mathrm{erf}}
\newcommand{\opt}{_{\mathrm{opt}}}
\begin{document}

\title{The role of a matchmaker in buyer-vendor interactions}
\author{Linyuan L\"u, Mat\'u\v s Medo, Yi-Cheng Zhang}
\authorrunning{Linyuan L\"u \emph{et al}}
\institute{Physics Department, University of Fribourg,
P\'erolles, 1700~Fribourg, Switzerland\and Lab of Information
Economy and Internet Research, University of Electronic Science
and Technology of China, 610054~Chengdu, China}

\abstract{We consider a simple market where a vendor offers
multiple variants of a certain product and preferences of both
the vendor and potential buyers are heterogeneous and possibly
even antagonistic. Optimization of the joint benefit of the
vendor and the buyers turns the toy market into a combinatorial
matching problem. We compare the optimal solutions found with
and without a matchmaker, examine the resulting inequality
between the market participants, and study the impact of
correlations on the system.
}

\maketitle

\section{Introduction}
The study of complex economic systems has attracted attention of
many physicists. They have contributed to the field with several
highly simplified, yet influential, models such as the minority
game~\cite{CMZ05}, percolation~\cite{CB00}, scaling models of
financial markets~\cite{MS02}, as well as with a set of useful
tools and insights~\cite{MS99,Sornette03,Buchanan04,Helbing08}.

Adopting a simplifying point of view characteristic for the
works mentioned above, in this paper we focus on the
interactions between consumers and producers. These interactions
represent a classical example of decision-making under
uncertainty~\cite{BRT88} where the limited information
available to the contract participants results in a~risk of
making a wrong decision. In the standard economic literature,
problems related to the interactions of consumers and producers 
are as diverse as the research of consumer
behavior~\cite{SCDDN01}, the question of trust~\cite{DC97}, the
economics of information~\cite{Stigler61,Stiglitz06}, and the
behavior of entire firms and industries~\cite{CaPe05,Sut91}.

This work is particularly motivated by the classical
stable-marriage problem~\cite{GS62,LZ03} in which $N$ men and
$N$ women all have their individual preferences and are to be
matched one-to-one. Almost inevitably, it's impossible to
satisfy everyone and hence stable matchings (where no one has
the possibility to exchange the assigned partner for a better
one) or the optimal matching (where the total satisfaction is
maximal) are of interest. The stable marriage problem has
implications in many economic and social systems. It can
represent matching of job seekers and employers or that of
lodgers and landlords; it is also a metaphor for problems in
logistics~\cite{HP01} and in online marketing~\cite{Bakos98}.

We study a situation where a certain product is available in
multiple variants and the preferences of both the buyer and the
vendor for each of the variants can be represented by numbers
(the higher the number, the more appreciated the variant). The
matching of the buyer and the vendor is then achieved by the
selection of a single variant to be delivered. In the given
framework, we first study outcomes achieved with the help of an
external matchmaker---an idealized agent supervising the market
and having perfect information about all the preferences. In
particular, we investigate the inequality between profits
enjoyed by the two involved parties and how correlations of the
preferences influence the system's behavior.

For comparison, we study two simple matchmaker-free models of
variant selection. In the first matchmaker-free model, the
vendor makes consecutive offers and the buyer decides whether to
accept an offer or not. Our results show that while this
approach results in a small decrease of the total satisfaction,
it considerably decreases the buyer-vendor inequality. In the
second matchmaker-free model, the buyer is searching for the
optimal variant by himself. We study the optimal number of
examined variants and show that under some conditions, this
number may be infinite: the buyer is tempted to search forever.
Numerical simulations of the studied models are in most cases
accompanied with approximate analytical results.

\section{Trading under matchmaker's supervision}
We assume that a given product is available in $N$ different
variants which can be prepared by a vendor and fulfill, to
a~greater or lesser degree, needs of a given buyer. The buyer's
utility from purchasing variant $\alpha$ is denoted by
$x_{\alpha}$ and the vendor's utility from providing this
variant is denoted by $y_{\alpha}$. The matchmaker optimizes the
joint benefit by maximizing the total utility
$u_{\alpha}(x_{\alpha},y_{\alpha})$. Obviously, the system's
behavior depends on the choice of the utility function and on
the nature of the utilities $x_{\alpha},y_{\alpha}$---we shall
study different settings in the following sections.

\subsection{Linear utility function}
The simplest form of the total utility is
\begin{equation}
\label{util-linear}
u_{\alpha}(x_{\alpha},y_{\alpha})=x_{\alpha}+y_{\alpha}
\end{equation}
where both utilities are merely summed with equal weights. In
addition, we assume that both $x_{\alpha}$ and $y_{\alpha}$ are
random variables drawn from the uniform distribution in the
range $[-1,1]$, $\UU(-1,1)$, and that they are uncorrelated. The
distribution $f(u_{\alpha})$ then has the tent-shaped form
\begin{equation}
\label{fu-one}
f(u_{\alpha})=\bigg\{
\begin{array}{ll}
(2+u_{\alpha})/4\qquad & u_{\alpha}\in[-2;0),\\
(2-u_{\alpha})/4\qquad & u_{\alpha}\in[0;2].
\end{array}
\end{equation}
The probability that a randomly selected variant has the total
utility greater than $u_{\alpha}$ is
$P(u_{\alpha}):=\int_{u_{\alpha}}^2 f(u')\,\mathrm{d} u'$. Since
utilities of different variants are mutually independent, the
largest utility $u_m:=\max_{\alpha=1}^N u_{\alpha}$ has the
distribution
\begin{equation}
\label{extreme-u}
g(u_m)=Nf(u_m)[1-P(u_m)]^{N-1}.
\end{equation}
Here the factor $N$ appears because any of $N$ variants can have
the largest utility, the factor $f(u_m)$ is the occurrence
probability of $u_m$, and the factor $[1-P(u_m)]^{N-1}$ is the
probability that the remaining $N-1$ variants have utilities
lower than $u_m$; Eq.~\req{extreme-u} is also known as the
extreme statistics of the random variable $u_m$. We study the
model by computing $\avg{u_m}$ (by $\avg{x}$ we denote the
average of $x$ over all possible realizations). Since
$P(u_{\alpha}<0)=1/2$ and $u_m<0$ only when all $N$ variants
have $u_{\alpha}<0$, it follows that $P(u_m<0)=2^{-N}$. Hence,
assuming $N\gg1$, we can confine our computation to $u_m>0$
where $f(u_m)=(2-u_m)/4$ and $P(u_m)=(2-u_m)^2/8$. When $N$ is
large, $P(\avg{u_m})$ is small and thus we can use the
approximation $1-P(u_m)\approx\exp[-P(u_m)]$. After replacing
the lower integration bound in $\avg{u_m}$ with $-\infty$ (this
is again justified by the negligible probability of $u_m<0$) we
obtain
\begin{equation}
\label{u_m-avg}
\avg{u_m}\approx\frac{2N}{N-1}-\sqrt{\frac{2\pi N^2}{(N-1)^3}}
\approx 2-\sqrt{\frac{2\pi}{N}}
\end{equation}
where we neglected terms of order $O(1/N)$ and higher. We see
that as $N$ increases, $\avg{u_m}$ rapidly approaches its upper
bound---the difference scales with $N^{-1/2}$. Numerical
computation of $\avg{u_m}$ shows that the relative error of
Eq.~\req{u_m-avg} decreases fast with $N$: it is less than
$1\%$ already for $N=17$.

Apart from the optimal total utility $u_m$, the inequality
between the vendor and the buyer is also of interest. If variant
$\beta$ maximizes $u_{\alpha}$, we say that the buyer-vendor
inequality is $\abs{x_{\beta}-y_{\beta}}$ and denote its
expected value by $\Delta$. To compute $\Delta$, one needs to
realize that for any given $u_{\alpha}$, the term
$\abs{x_{\alpha}-y_{\alpha}}$ ranges from $0$ to $2-u_{\alpha}$
(this maximal difference is achieved when one of the two
utilities is $1$ and the other is $u_{\alpha}-1$). Since
$x_{\alpha}$ and $y_{\alpha}$ are uniformly distributed, all
possible values of $\abs{x_{\alpha}-y_{\alpha}}$ are equally
probable and hence
$\avg{\abs{x_{\alpha}-y_{\alpha}}}=1-u_{\alpha}/2$. In turn,
$\Delta=1-\avg{u_m}/2=\sqrt{N/(2\pi)}$ which can be easily
confirmed by numerical computation.

\subsection{Affecting the inequality}
Blind maximization of the total utility may not be the best
policy because it can result in large inequalities between
society members. In an effort to prevent that, the matchmaker
may adopt the utility function
\begin{equation}
\label{util-nonlinear}
u_{\alpha}'(x_{\alpha},y_{\alpha})=
\big(x_{\alpha}^k+y_{\alpha}^k\big)^{1/k}\qquad (k>0).
\end{equation}
where the variants that have one or both utilities negative
are automatically excluded (they do not have the chance to
become selected anyway). Choosing $k\gg1$ in
Eq.~\req{util-nonlinear} favors those pairs
$(x_{\alpha},y_{\alpha})$ where at least one of the utilities is
high, while $k<1$ favors more equal splitting of the total
utility. The expected value of
$u_m':=\max_{\alpha=1}^N u_{\alpha}'$ can be computed in the
same way as in the previous section, yielding
\begin{equation}
\label{u_m-general}
\avg{{u_m}'}\approx
2-\left(\frac{\Gamma(\tfrac12+\tfrac1k)\sqrt{\pi}}
{\Gamma(1+\tfrac1k)4^{1-1/k}}\right)^{1/2}N^{-1/2}.
\end{equation}
Interestingly, $\avg{u_m'}$ scales with $N$ in the same way as
$\avg{u_m}$.

\begin{figure}
\centering
\includegraphics[scale=0.3]{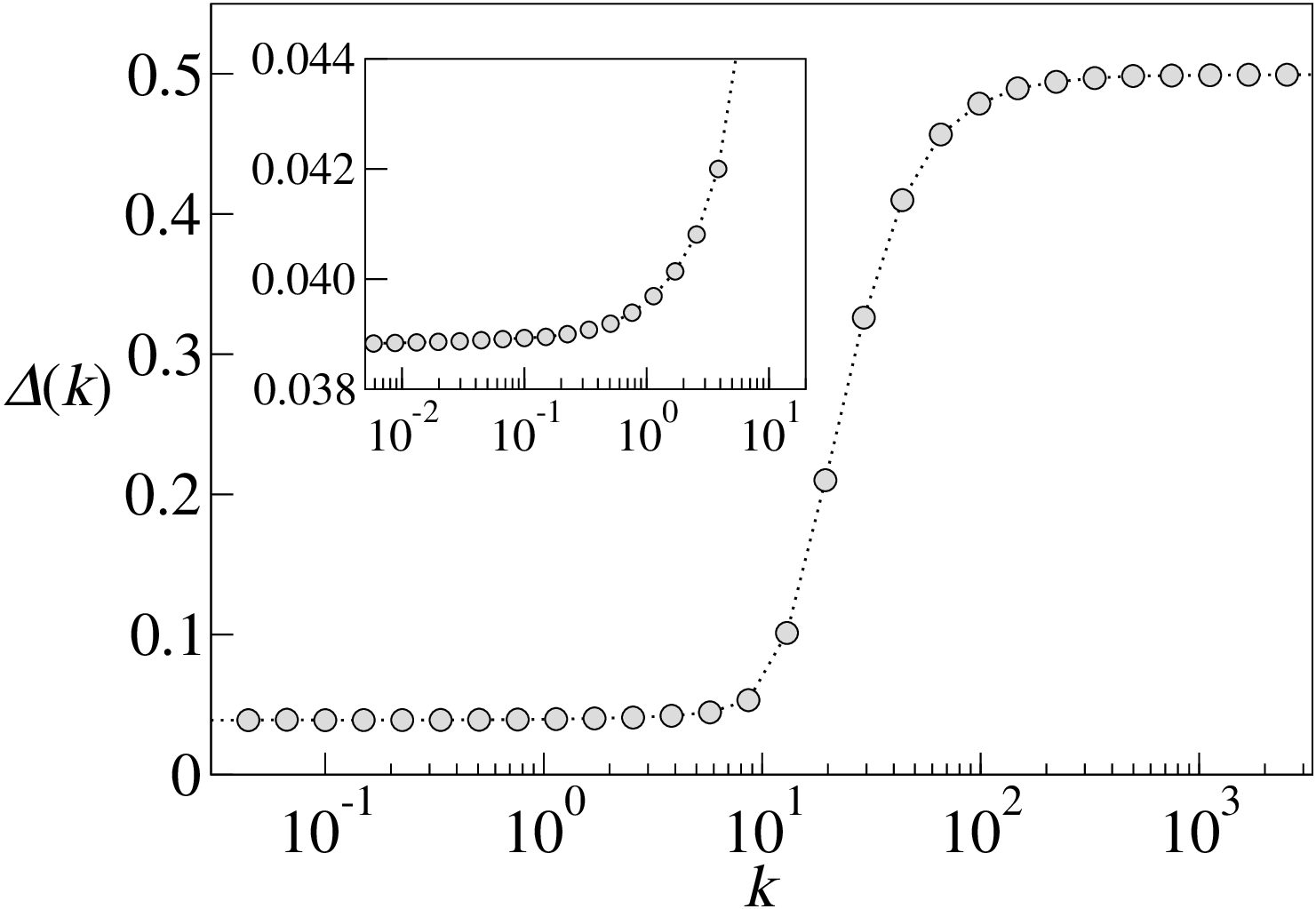}
\caption{Inequality $\Delta(k)$ as a function of $k$ for
$N=1\,000$. The inset focuses on small values of $k$ and the
results are averaged over $10\,000$ independent realizations.}
\label{fig:inequality}
\end{figure}

The expected buyer-vendor inequality is now a function of $k$,
$\Delta(k)$. Numerical results shown in
Fig.~\ref{fig:inequality} confirm our initial insight that
$\Delta(k)$ grows with $k$. In particular, the value $0.5$
achieved for $k\gtrsim100$ corresponds to picking the variant
which maximizes one of the utilities and paying no attention to
the other utility. In that case, the selected variant has the
larger of the two utilities close to $1$ and the other one is
$0.5$ on average: together we have
$\lim_{k\to\infty}\Delta(k)=0.5$. Further, it can be seen in
Fig.~\ref{fig:inequality} that $k<1$ does not significantly
decrease the inequality $\Delta(k)$ and thus to create a more
social society, one has to use a different utility function. For
example, for $N=1\,000$, optimizing the outcome of the weaker
(which corresponds to $u_{\alpha}(x_{\alpha},y_{\alpha})=
\min[x_{\alpha},y_{\alpha}]$) decreases the inequality by $29\%$
while reducing the total utility by less than $0.3\%$.

\subsection{Serving several buyers at once}
\begin{figure}
\centering
\includegraphics[scale=0.3]{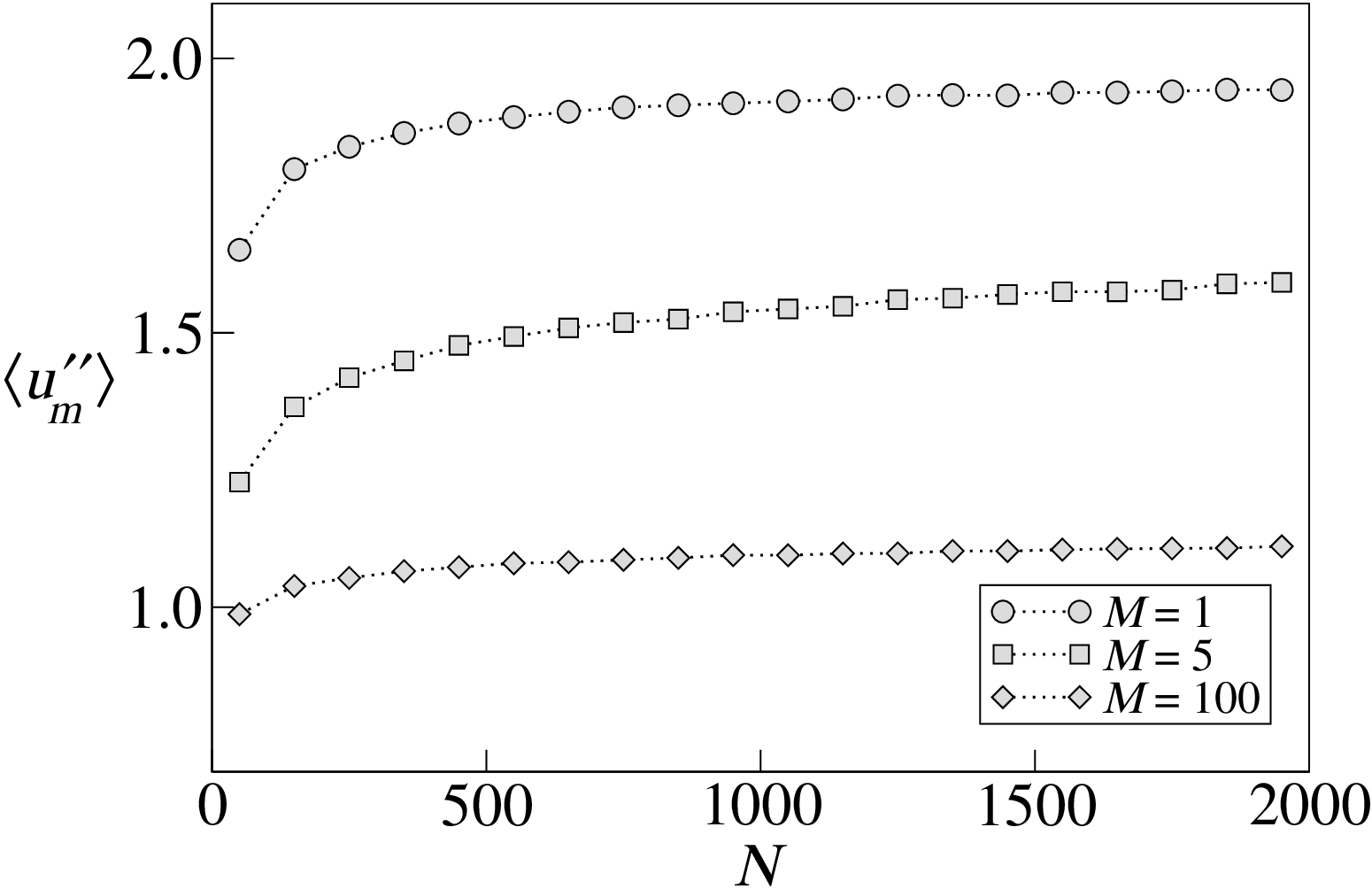}
\caption{Average utility with multiple buyers (results are
averaged over $1\,000$ independent realizations).}
\label{fig:Mbuyers}
\end{figure}

When there are several buyers in the market, the matchmaker can
either find the best variant for each buyer separately or she
can compromise buyers' needs by finding one variant for all.
While the former case is identical with our analysis above, the
latter case is different and requires the straightforward
generalization of the total utility to the form
$My_{\alpha}+\sum_{i=1}^M x_{i,\alpha}$ where $M$ is the number
of buyers and $x_{i,\alpha}$ is the utility of variant $\alpha$
for buyer $i$. For consistency with our previous formalism, we
introduce the per-buyer utility
\begin{equation}
\label{util-Mbuyers}
u_\alpha''(x_{1,\alpha},\dots,x_{M,\alpha},y_{\alpha})=
y_{\alpha}+\frac1M\sum_{i=1}^M x_{i,\alpha}:=
y_{\alpha}+a_{\alpha}
\end{equation}
where $a_{\alpha}$ is the average buyers' utility of object
$\alpha$, the maximal utility we denote as $u_m''$. When
utilities $x_{i,\alpha}$ are independent and the number of
buyers is large, the central limit theorem states that the
difference $a_{\alpha}-\avg{x_{i,\alpha}}$ is approximately
a~normally distributed quantity with variance proportional to
$1/M$. It follows that due to the fast decay of the normal
distribution, the matchmaker cannot find an object with the
average utility $a_{\alpha}$ differing substantially from
$\avg{x_{i,\alpha}}$. This is confirmed by
Fig.~\ref{fig:Mbuyers} where we show $\avg{u_m''}$ for various
values of $M$. As $M$ increases, fluctuations of $a_{\alpha}$
gets smaller and $\avg{u_m''}\approx1$, corresponding to
a~negligible contribution of $a_{\alpha}$ to $u_{\alpha}''$ (it
can be shown that to achieve non-negligible
$\max_{\alpha} a_{\alpha}$, the necessary number of variants is
proportional to $\ee^M$).

Now we see that it's impossible to satisfy several buyers with
one variant. Since production of an individual variant for each
buyer is often too expensive, it is then a natural question how
to compromise between the buyers' satisfaction and the costs of
personalized production. Within the given framework, one can
introduce an additional cost which increases with the number of
variants produced by the vendor---such a cost forces the vendor
to narrow down the selection. This aspect of buyer-vendor
interactions is extensively studied in~\cite{Medo08,Lu08} where
they show that based on the compromise described above and a few
simple additional assumptions, one can reproduce a rich variety
of market phenomena.

\subsection{Correlated utilities}
So far we assumed that the vendor's and buyer's utilities are
mutually uncorrelated. While convenient for analytical
computation, this is not a realistic assumption because in
general: what is good for the vendor is not good for the buyer
and vice versa. In other words, one expects $x_{\alpha}$ and
$y_{\alpha}$ to be negatively correlated.

To study the influence of correlations we first need to find
a~way how to introduce them into the system. In general it is
easy to create correlated quantities by introducing a control
parameter $t\in[0,1]$ and assuming that both quantities have
a~common part which is proportional to $t$ and independent parts
which are proportional to $1-t$. However, when each part itself
is uniformly distributed, the resulting distribution depends on
the value of $t$ and this effect distorts further analysis of
the system~\cite{Medo08}. This motivates us to switch from
uniform distributions to normal distributions which preserve
their functional form under addition. We assume that utilities
$x_{\alpha}$ and $y_{\alpha}$ are obtained as
\begin{equation}
\label{util-correlated}
x_{\alpha}=\sqrt{1-t}\,X_{\alpha}+\sqrt{t}\,C_{\alpha},\quad
y_{\alpha}=\sqrt{1-t}\,Y_{\alpha}+s\sqrt{t}\,C_{\alpha}
\end{equation}
where $X_{\alpha},Y_{\alpha},C_{\alpha}$ are drawn from
the standard normal distribution $\NN(0,1)$, $t\in[0,1]$ is the
parameter controlling the correlation strength and the parameter
$s$ switches between positive ($s=1$) and negative ($s=-1$)
correlations. Since when adding two normal distributions,
individual means and individual variances sum up to give the
resulting mean and variance respectively, both $x_{\alpha}$ and
$y_{\alpha}$ have zero mean and unit variance. It is simple to
compute the Pearson correlation coefficient of $x_{\alpha}$ and
$y_{\alpha}$ which is $C_{xy}=st$.

\begin{figure}
\centering
\includegraphics[scale=0.3]{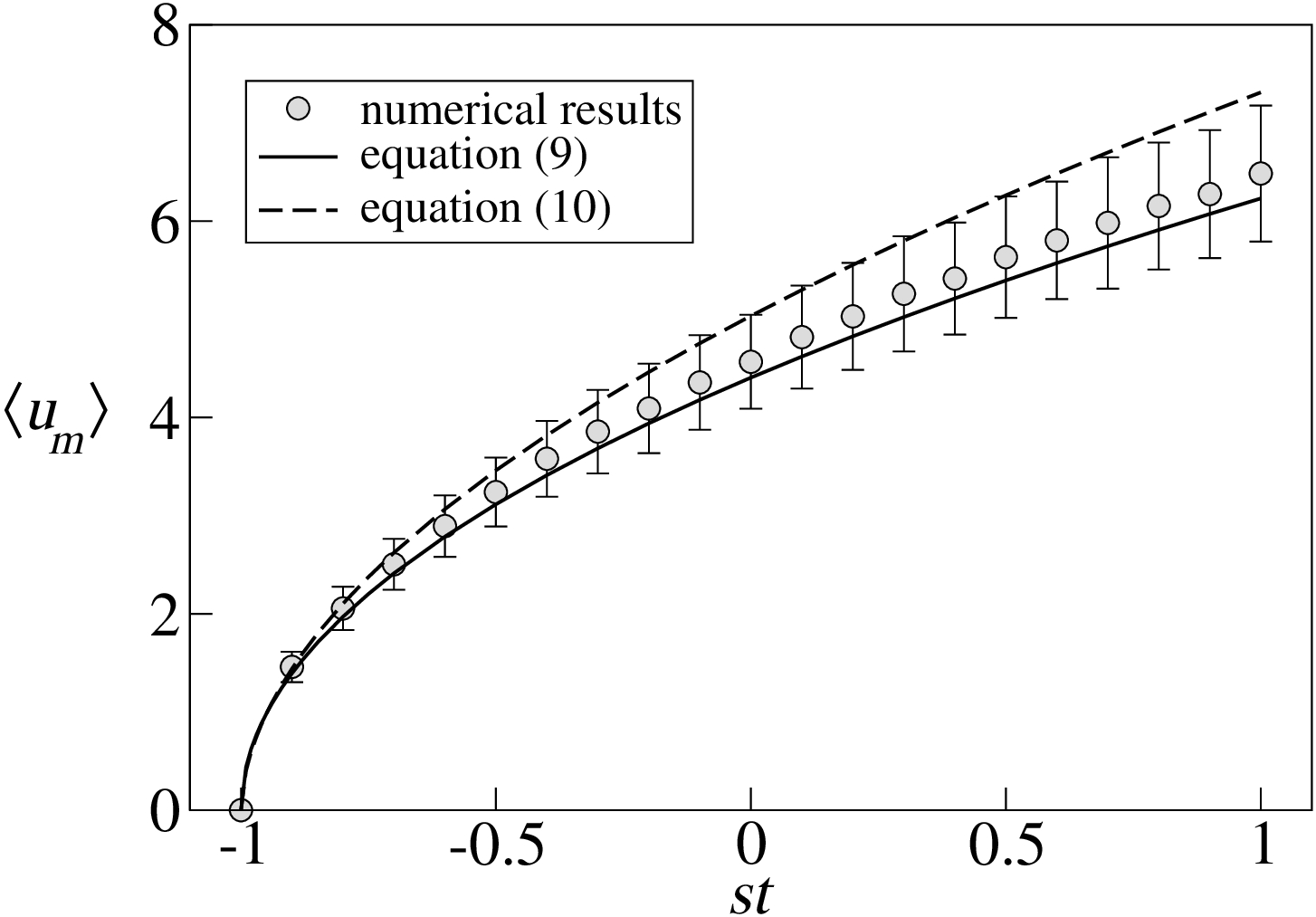}
\caption{The dependence of the average maximal utility
$\avg{u_m}$ on $st$ for $N=1\,000$ (numerical results and their
standard deviations are obtained from 1\,000 realizations of the
model).}
\label{fig:gaussian}
\end{figure}

To study the system of utilities produced by
Eq.~\req{util-correlated}, we assume the linear total utility
given by Eq.~\req{util-linear}. Hence $u_{\alpha}$ is normally
distributed with zero mean and its variance can be shown to be
equal to $2(1+st):=v$. We are again interested in
$u_m:=\max_{\alpha=1}^N u_{\alpha}$ and $\avg{u_m}$. It can be
shown (see Appendix~\ref{appendix1} for details) that
$\avg{u_m}$ approximately solves the equation
\begin{equation}
\label{gauss-eq}
\avg{u_m}\exp\bigg[\frac{\avg{u_m}^2}{4(1+st)}\bigg]=
N\sqrt{\frac{1+st}{\pi}}.
\end{equation}
A comparison of this result with a numerical computation of
$\avg{u_m}$ (where we randomly generate the utilities
$x_{\alpha}$ and $y_{\alpha}$, find the maximal total utility
$u_m$ and average over many realizations) is shown in
Fig.~\ref{fig:gaussian}. As we can see, positive correlations
amplify the variance of $u_{\alpha}$ and hence allow the
matchmaker to reach a higher optimal utility. Notice that by
setting $t=0$ in Eq.~\req{gauss-eq}, one automatically obtains
the result for normally distributed uncorrelated utilities.

More insight can be gained if we attempt to find an approximate
solution of Eq.~\req{gauss-eq}. When $N$ is large, the factor
$\avg{u_m}$ on the left side of Eq.~\req{gauss-eq} is much
smaller than the exponential term and hence it can be neglected.
The simplified equation can be solved and gives us the
approximate result
\begin{equation}
\label{gauss-1stapprox}
\avg{u_m}^2\approx 4(1+st)\ln\big[N\sqrt{(1+st)/\pi}\,\big].
\end{equation}
In contrast to Eq.~\req{u_m-avg}, this time $\avg{u_m}$ grows
with $N$ without bounds. On the other hand, this growth is
extremely slow: $\avg{u_m}$ is proportional to the square root
of $\ln N$. For example, in the uncorrelated case increasing $N$
from $1\,000$ to $1\,000\,000$ increases $\avg{u_m}$ only by
$50\%$.

\section{Trading without the matchmaker}
Despite all the results obtained so far, one question remains
open: what is the matchmaker's contribution to the studied
vendor-buyer matchings? This question can be answered by
investigating matchmaker-free methods of the variant selection.

\subsection{Vendor proposes}
In~\cite{Medo08} they assumed that when the vendor offers a
variant, the buyer accepts it only if his cost is smaller than
the vendor's cost (instead of maximization of utilities, they
studied minimization of costs). In our framework this means that
the buyer accepts variant $\alpha$ only if
$x_{\alpha}\geq y_{\alpha}$ (\emph{i.e.}, he wants to profit
more than the vendor). The vendor's advantage is that he decides
which variants to propose---obviously, it is optimal to begin
with the variant that maximizes $y_{\alpha}$. This concept,
where proposing and accepting sides are well defined and
distinguished, is similar to the classical Gale-Shapley
algorithm known from the stable marriage problem~\cite{GS62}.

\begin{figure}
\centering
\includegraphics[scale=0.3]{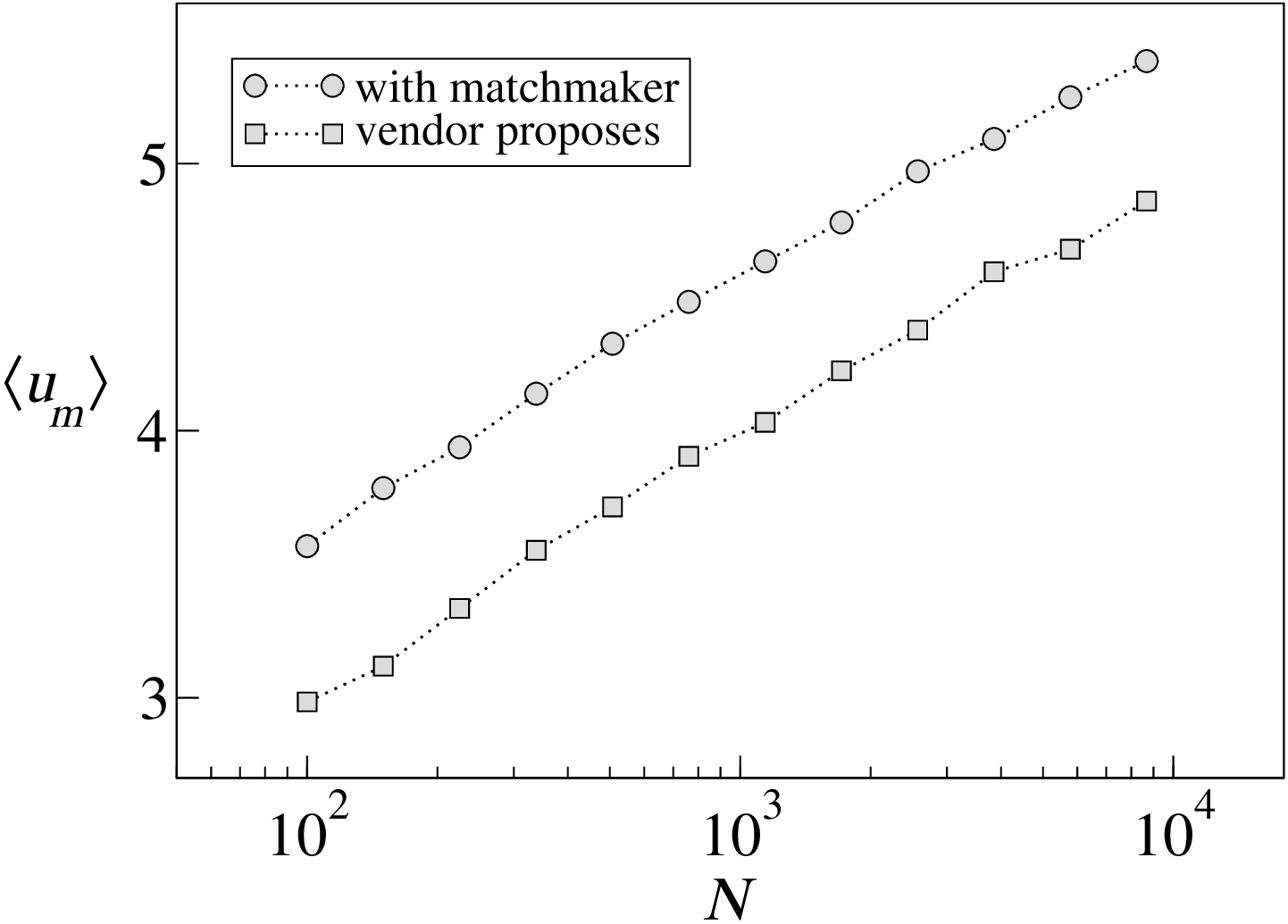}
\caption{The total utility of the selected variant with and
without the matchmaker. Individual utilities are drawn from
$\NN(0,1)$, results are averaged over $1\,000$ realizations.}
\label{fig:lin}
\end{figure}

Based on the matching described above and using
Eq.~\req{util-linear}, we can compute the average total utility
of the selected variant and compare it to the case with the
matchmaker. Assuming normally distributed utilities and zero
correlations, we studied the system numerically. As can be seen
in Fig.~\ref{fig:lin}, the total utility with the matchmaker is
considerably higher. On the other hand, the inequality between
the fellows decreases from approximately $1.1$ (with the
matchmaker) to approximately $0.4$ (when the vendor proposes).
When correlations are present, the difference between the two
matching methods decreases and becomes zero for $C_{xy}=\pm1$.

\subsection{Buyer's search}
As we have seen, the matchmaker optimizes the total utility at
the cost of compromising the utilities of individuals. The buyer
can avoid being ``compromised'' by searching for the best
variant by himself. The drawback is that the search is costly
(it consumes buyer's time and attention) and the corresponding
cost has to be subtracted from the utility of the eventually
selected variant. Since the time spent by searching grows
linearly with the number of examined variants, it is natural to
assume the linear cost term $\beta N$, where $N$ is the number
of examined variants and $\beta>0$ is the cost per examined
variant. The expected buyer's utility is
\begin{equation}
\label{util-search}
u_S(\beta,N)=\big\langle\max_{1\leq\alpha\leq N}
x_{\alpha}\big\rangle-\beta N:=\avg{x_m}-\beta N.
\end{equation}
This form of the buyer's utility was suggested
in~\cite{Stigler61} where, however, the emphasis was on the
discussion of the cost of information. In today's computerized
and networked world, searching is easy. Hence to obtain
approximate analytical results, we assume that $\beta$ is much
smaller than the typical utility value.

\begin{figure}
\centering
\includegraphics[scale=0.3]{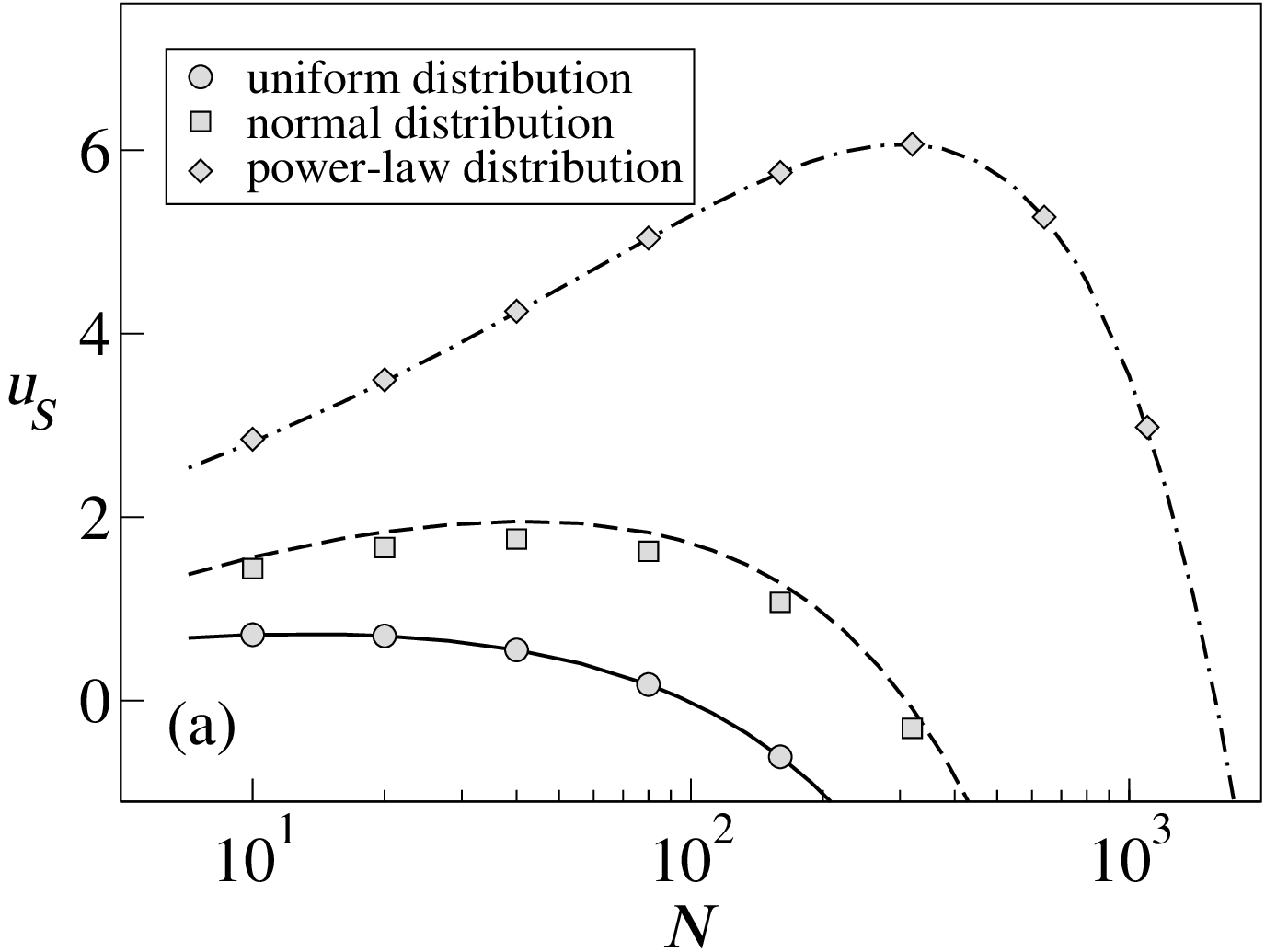}\\[6pt]
\includegraphics[scale=0.3]{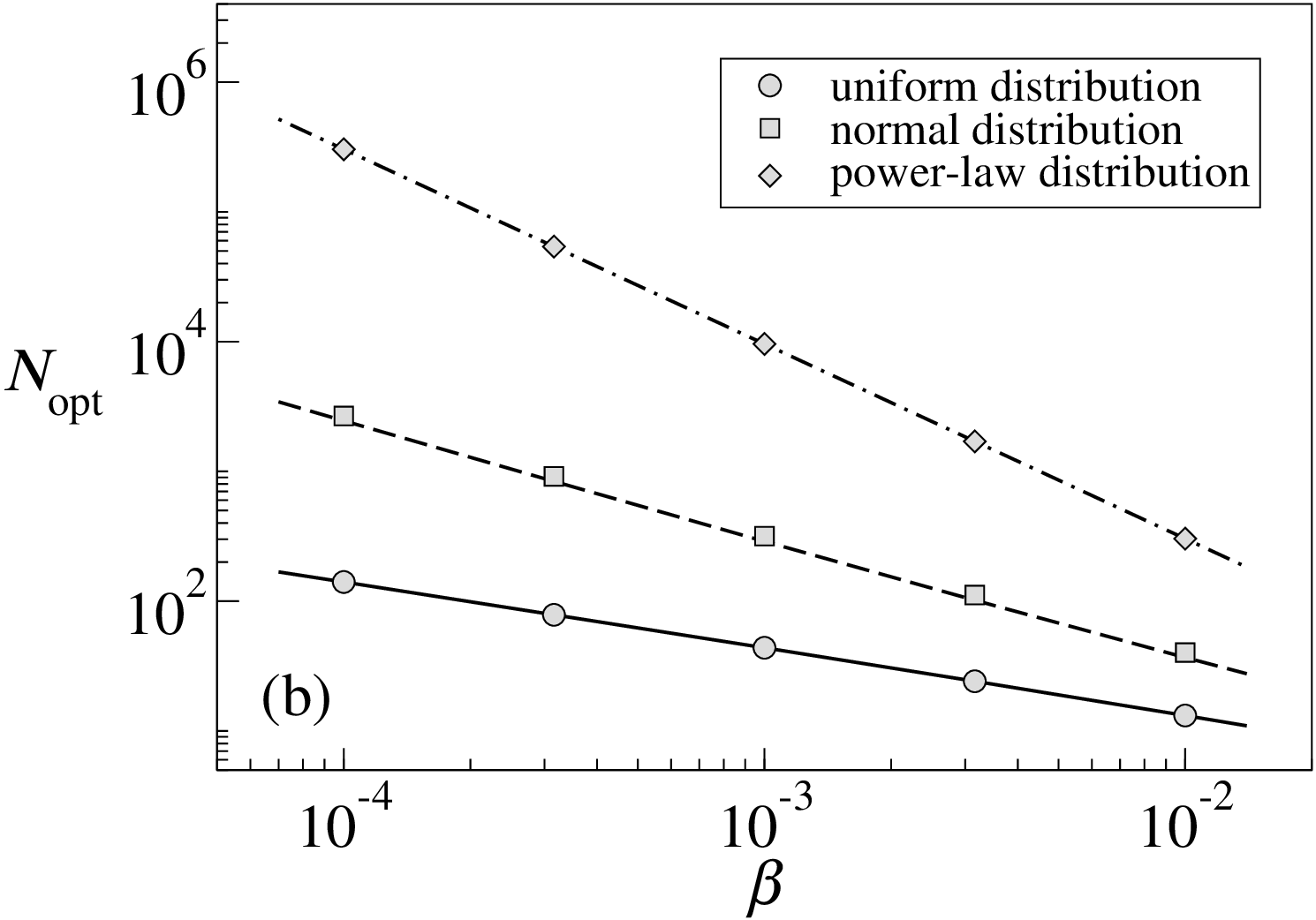}
\caption{The buyer's search for three distinct probabilistic
distributions of utilities---comparison of numerical results
(shown with symbols) and analytical results (shown with lines)
for $u_S(0.01,N)$ (a) and $N\opt$ (b); $\gamma=4$ in both
figures.}
\label{fig:search}
\end{figure}

In Eq.~\req{util-search}, the term $\avg{x_m}$ grows with $N$
but the cost $\beta N$ eventually takes over and the total
utility $u_S(\beta,N)$ decreases (see Fig.~\ref{fig:search}a for
an illustration). This behavior is in agreement with the
classical observation ``Good things satiate, bad things
escalate.'' by psychologists Coombs and Avrunin~\cite{CooAvr77}.
It is now natural to ask, what number of examined variants
$N\opt$ maximizes $u_S(\beta,N)$. We shall do that in three
distinct cases. In addition to $x_{\alpha}$ drawn from the
uniform distribution $\UU(-1,1)$ and from the normal
distribution $\NN(0,1)$, which were studied before, we consider
also the case when $x_{\alpha}$ is drawn from the power-law
distribution $f(x)=(\gamma-1)x^{-\gamma}$, $x\in[1,\infty)$,
$\gamma>2$ (for accounts on the importance of power-law
distributions in complexity management and organization see
\emph{e.g.}~\cite{Buchanan04,McKeAn05}).

In Appendix~\ref{appendix2} we find expressions for $\avg{x_m}$
in all three cases. Therein, the optimal number of variants
$N\opt$ is found in the forms
\begin{eqnarray}
\label{Nopt-uniform}
\mbox{uniform: }&&N\opt=
(\beta/2)^{-1/2}-1,\\
\label{Nopt-gaussian}
\mbox{normal: }&&N\opt\approx
\big(\beta\sqrt{-\ln(2\pi\beta^2)}\big)^{-1}, \\
\label{Nopt-powerlaw}
\mbox{power-law: }&&N\opt\approx
\big(\beta(\gamma-1)/\Gamma(\delta)\big)^{-1/\delta},
\end{eqnarray}
where $\delta=(\gamma-2)/(\gamma-1)$. Noticeably, in all three
cases we observe a power-law dependency of $N\opt$ on $\beta$.
In the case of uniformly distributed utilities, $\avg{x_m}$ has
an upper bound and with $N$ it grows very slowly (that means:
there is little to be gained by an extensive search). In
consequence, $N\opt$ is proportional to $\beta^{-1/2}$ and hence
it is little sensitive to changes of $\beta$. When utilities are
normally distributed, $\avg{x_m}$ grows with $N$ without
bounds---this allows $N\opt$ to reach higher values than in the
former case. In the power-law case, values of $N\opt$ are largest
and their dependency on $\beta$ is strongest. Moreover, as the
power-law exponent $\gamma$ gets closer to $2$, the growth of
$\avg{x_m}$ with $N$ becomes linear and hence when the growth
rate is larger than $\beta$, it is optimal for the buyer to
search ``forever'' (the exponent $\delta$ diverges).

To review the accuracy of the presented results, in
Fig.~\ref{fig:search} we compare them with numerical
simulations. As can be seen, good agreement is achieved in all
three cases. Let us conclude with the remark that in the
power-law case, the average, the mode, and the median of $x_m$
may differ significantly. In consequence, it is an important
question whether the buyer should rely on $\avg{x_m}$ which is
strongly affected by rare extreme events.

\section{Conclusion}
In this paper, we studied several simplified models of
interactions between buyers and a vendor. Assuming that both the
vendor and the buyers can attribute certain utility values to
each available variant, the decision which variant is to be
delivered can be formulated as a mathematical optimization
problem.

Firstly we assumed that there is a matchmaker who can fairly
select the optimal variant according to what is best for all
participants of the contract. A plausible criterion for the
matchmaker is to select the variant with the maximal total
utility. Our results show that when the utilities are uniformly
distributed, the matchmaker can achieve the total utility close
to its upper bound even when the number of available variants is
small. In other words, little choice is enough in this case. On
the other hand, when the total utility is maximized, the
difference between utilities of the vendor and the buyer may be
too large to consider it to be the optimal choice. When the mere
summation of individual utilities is replaced by a more refined
expression, the inequality between the involved parties may be
decreased. In particular, we found that when the matchmaker
maximizes the smaller of the two individual utilities, the
inequality decreases significantly while the total utility is
almost unchanged.

When the matchmaker tries to find one variant for several
buyers, the situation turns out to be far less favorable: the
number of variants needed to approach the upper bound of the
total utility grows exponentially with the number of buyers. As
a result, production of multiple variants is advisable---this
topic was extensively studied in our previous
works~\cite{Medo08,Lu08}. Assuming normally distributed
utilities, we studied the influence of correlations on the
system behavior. While analytically more demanding, this
generalization is important because it makes the system more
realistic. Our results confirm that the optimal total utility
depends strongly on the correlation of utilities.

Secondly we studied the variant selection without a matchmaker.
In this case, the space of possible means of variant selection
is vast and hence we focused on two particular situations. In
the first one, the vendor offers and the buyer passively decides
whether to accept the offered variant or not. While in
comparison with the matchmaker-mediated outcome, the total
utility is slightly lower (see Fig.~\ref{fig:lin}), the
inequality is decreased substantially. In the second selection
method, the buyer chooses the variant by himself but he also has
to pay the cost for examining the variants. Here we considered
three different distributions of utilities and shown that when
the distribution has a power-law tail, for the buyer it may be
optimal to inspect a huge number of variants. In the extreme
case of a power-law distribution with the exponent lower or
equal than $2$, the buyer does best by searching forever.
Admittedly, these results are influenced by the linear growth of
searching cost with the number of examined variants $N$ assumed
by Eq.~\req{util-search}. In contrast, in psychology it is well
known that refusing the second best variant or having too many
options may be frustrating for people~\cite{Schw03}. These
effects could be included by adding an additional cost term
depending on the utility of the second best variant, or simply
by a part of the searching cost proportional to a higher power
of $N$.

\appendix

\section{Extreme statistics for normally distributed variables}
\label{appendix1}
In this appendix we show how $\avg{u_m}$ can be approximated
when the number of variants, $N$, is large. Assuming
$u_{\alpha}\in\NN(0,v)$, $u_m:=\max_{\alpha=1}^N u_{\alpha}$
has the distribution
\begin{equation}
\label{start}
f(u_m)=\frac{N\ee^{-u_m^2/2v}}{\sqrt{2\pi v}}
\left(1-\int_{u_m}^{\infty}
\frac{\ee^{-u^2/2v}}{\sqrt{2\pi v}}\dd u\right)^{N-1}.
\end{equation}
The error function
$\erf[x]:=\frac2{\sqrt{\pi}}\int_0^x\exp[-t^2]\,\dd t$
allows us to write the integral in Eq.~\req{start} as
$\tfrac12-\erf[u_m/\sqrt{2v}]/2$. When $N$ is large,
$u_m^2/(2v)\gg1$ and hence we can use the asymptotic
expansion $\erf[x]\approx1-\exp[-x^2]/(x\sqrt{\pi})$
(taken from \texttt{mathworld.wolfram.com}, the next
contributing term is proportional to $\exp[-x^2]/x^3$). Again,
we use the approximation $1-y\approx\exp[-y]$ (which is valid
for $y\ll1$) to obtain
$$
f(u_m)\approx\frac{N}{\sqrt{2\pi v}}\exp\Big[-\frac{u_m^2}{2v}-
\frac{(N-1)\exp[-u_m^2/2v]}{u_m\sqrt{2\pi/v}}\Big].
$$
Unfortunately, the integral corresponding to $\avg{u_m}$ cannot
be solved. Since $f(u_m)$ does not have heavy tails,
a~reasonably precise result can be obtained by approximating
$\avg{u_m}$ with $\tilde u_m$ maximizing $f(u_m)$, yielding
$$
\avg{u_m}=(N-1)(1+v/\avg{u_m}^2)
\sqrt{\frac{v}{2\pi}}\exp[-\avg{u_m}^2/2v].
$$
When $N$ is large, $v/\avg{u_m}^2\ll1$ and hence we neglect this
term on the right side. In addition, $N-1\approx N$ and we get
\begin{equation}
\label{gauss-general-result}
\avg{u_m}\exp\big[\avg{u_m}^2/2v\big]=N\,\sqrt{v/2\pi}
\end{equation}
which after substituting $v=2(1+st)$ gives Eq.~\req{gauss-eq}.

\section{Analysis of the buyer's search}
\label{appendix2}
We begin with $x_{\alpha}\in\UU(-1,1)$. Then $x_m$ has the
distribution $g(x_m)=\tfrac{N}{2}\big[1-(1-x_m)/2\big]^{N-1}$
and in consequence $\avg{x_m}=1-2/(N+1)$. Maximizing
$u_S(\beta,N)$ with respect to $N$ we get
$N\opt=\sqrt{2/\beta}-1$.

Now let's consider $x_{\alpha}\in\NN(0,1)$. After substituting
$v=1$ in Eq.~\req{gauss-general-result}, the expected utility
$u_S(\beta,N)$ can be maximized using implicit derivative,
yielding the condition $\beta N\avg{x_m}=1$. In this equation,
``fast'' and ``slow'' terms ($N$ and $\avg{x_m}$ respectively)
are mixed together and hence an approximate solution can be
found by the following iterative procedure. Approximating
$\avg{x_m}=1$ gives the rough estimate $N_0=1/\beta$. Together
with Eq.~\req{gauss-general-result}, this value leads to the
improved estimate $\avg{x_m}\approx\sqrt{-\ln(2\pi\beta^2)}$ and
in turn we get
$$
N\opt\approx\big(\beta\sqrt{-\ln(2\pi\beta^2)}\big)^{-1},
$$
an improved estimate of $N\opt$.

Finally let's study the case where utilities $x_{\alpha}$ are
constrained to the range $[1,\infty)$ and follow the power-law
distribution $f(x_{\alpha})=(\gamma-1)x^{-\gamma}$, $\gamma>2$.
Then the cumulative distribution is $P(x_{\alpha})=x^{1-\gamma}$
and the distribution of $x_m$ is
$g(x_m)=N(\gamma-1)x^{-\gamma}(1-x^{1-\gamma})^{N-1}$. After
approximating $1-x_m^{1-\gamma}\approx\exp[-x_m^{1-\gamma}]$
(the relevant values of $x_m$ are large) and replacing the lower
integration bound with $0$ we obtain
$\avg{x_m}\approx N^{1/(\gamma-1)}\Gamma[\delta]$ where
$\delta=(\gamma-2)/(\gamma-1)$; this result is well defined for
all $\gamma>2$. When only a fraction $r$ of all utilities
follows the power law, the result generalizes to
$\avg{x_m}\approx(Nr)^{1/(\gamma-1)}\Gamma[\delta]$.
Maximization of the expected utility $u_S(\beta,N)$ is
straightforward and yields $N\opt\approx
\big[\beta(\gamma-1)/\Gamma[\delta]\big]^{-1/\delta}$. For
a~different analysis of the largest value of a power-law
distributed sample and a~broader discussion of power laws
see~\cite{Newm05}.

\end{document}